\documentclass[prd,aps,preprintnumbers,preprint,tightenlines,amsmath,amssymb,showpacs,superscriptaddress,longbibliography]{revtex4-1}

\usepackage[T1]{fontenc}
\usepackage[latin9]{inputenc}
\usepackage{float}
\usepackage{bm}
\usepackage{dcolumn}
\usepackage{color}
\usepackage{amsmath}
\usepackage{amssymb}
\usepackage{graphicx} 
\usepackage{hyperref}
\hypersetup{colorlinks,citecolor=blue}
\usepackage[normalem]{ulem}

\graphicspath{{Pictures/}}

\begin{document}

\title{Material matter effects in gravitational UV/IR mixing} 

\author{Joseph Bramante} 
\affiliation{The McDonald Institute and Department of Physics, Engineering Physics, and Astronomy, Queen's University, Kingston, Ontario, K7L 2S8, Canada}
\affiliation{Perimeter Institute for Theoretical Physics, Waterloo, Ontario, N2L 2Y5, Canada}
\author{Elizabeth Gould}
\affiliation{The McDonald Institute and Department of Physics, Engineering Physics, and Astronomy, Queen's University, Kingston, Ontario, K7L 2S8, Canada}
\affiliation{Perimeter Institute for Theoretical Physics, Waterloo, Ontario, N2L 2Y5, Canada}

\date{\today}

\begin{abstract}
We propose a matter effect for the gravitational ultraviolet/infrared (UV/IR) mixing solution to the cosmological constant problem. Previously, the gravitational UV/IR mixing model implied a non-standard equation of state for dark energy, contradicting observation. In contrast, matter effect gravitational UV/IR mixing accommodates a standard $\Lambda$CDM cosmology with constant dark energy. Notably, there are new density-dependent predictions for futuristically precise measurements of fundamental parameters, like the magnetic moments of the muon and electron.
\end{abstract}

\maketitle

\section{Introduction}
\label{sec:intro}

The cosmological constant problem can be expressed as follows. Let us compare the observed vacuum energy in our universe $V_{obs}^4 \sim (2~{\rm meV})^4$ to the zero-point loop vacuum energy contribution from a field with mass $m$ and momentum $p$,
\begin{align}
\delta \rho_\Lambda \simeq \int_{\Lambda_{IR}}^{\Lambda_{UV}}  \frac{dp^3}{(2 \pi)^3} \frac{1}{2} \sqrt{p^2 + m^2} \simeq \frac{\Lambda_{UV}^4}{16 \pi^2},
\label{eq:cc}
\end{align}
where the last expression has assumed $m \ll \Lambda_{UV}$ and for now we neglect the infrared cutoff, setting $\Lambda_{IR}=0$. Standard Model field dynamics have been explored up to an ultraviolet cutoff $\Lambda_{UV} \sim ~{\rm TeV}$. With no evidence for a mechanism that cancels these contributions to the cosmological constant, after probing energies up to a TeV at colliders, the straightforward prediction for vacuum energy from field fluctuations in our universe is at least $\sim{\rm TeV^4}$, which is nearly sixty orders of magnitude larger than observed. For a review of the cosmological constant problem, see \cite{Weinberg:1988cp}. 

A crucial assumption in the standard cosmological constant argument is that the local quantum field theory underlying the loop calculation in Eq.~\eqref{eq:cc} is validly applied up to a cutoff $\Lambda_{UV}$ across a region the size of our universe. The gravitational UV/IR mixing solution to the cosmological constant problem proposed by Cohen, Kaplan, and Nelson \cite{Cohen:1998zx} (CKN) argues to the contrary, that effective field theory breaks down for any UV cutoff $\Lambda_{UV} \gtrsim {\rm meV}$ in a region the size of our universe.

The gravitational UV/IR mixing proposal follows from the observation that there will be inherently nonlocal gravitational dynamics in sufficiently large and dense systems. If loop-level zero-point field energy densities over a space $L$ are so large that a black hole forms inside $L$, we have entered a computational regime for which our local effective field theory has broken down, unless the theory properly accounts for the presence of virtual black hole states in loop-level field dynamics. Put another way, we should expect that the infrared cutoff of our theory $\Lambda_{IR} = 1/L$, implies a UV cutoff at around the threshold for black hole formation. It follows that zero-point loop corrections to vacuum energy in a properly defined effective field theory can only be reliably calculated if
\begin{align}
L \gtrsim R_{S} = 2G \left(\delta \rho_\Lambda \right) \frac{4 \pi}{3} L^3,
\end{align}
where $\frac{4 \pi}{3} L^3$ is the volume of the space, $ R_{S} = 2G M$ is the black hole horizon, $\delta \rho_\Lambda$ is the loop contribution to the vacuum energy density, and $G = \frac{1}{M_P^2}$ is Newton's constant.
Saturating this inequality leads to a maximum vacuum energy contribution from loop corrections,
\begin{align}
\delta \rho_\Lambda &\lesssim \frac{3}{8 \pi} \frac{M_{P}^2 }{ L^2}.
\label{eq:ineqorig}
\end{align}
In essence, this inequality reflects that Eq.~\eqref{eq:cc} should only be evaluated up to a UV cutoff $\Lambda_{UV} = \delta \rho_\Lambda^{1/4} \simeq \sqrt{M_{P} / L} $, consistent with our ignorance of quantum mechanics around black holes, whose size is determined by the IR cutoff $\Lambda_{IR} = 1/L$.

If the Hubble parameter is used as an IR cutoff for the universe \citep{Cohen:1998zx}, $\Lambda_{IR} = 1/L = H$, the prediction for loop contributions to vacuum energy is $\delta \rho_\Lambda \sim {\rm meV^4}$, which is around the size of the observed vacuum energy. This can be verified by comparing Eq.~\eqref{eq:ineqorig} to the reduced Friedmann equation for a background density $\rho$ in flat space,
\begin{align}
\frac{3 }{8 \pi} H^2 M_P^2= \rho ,
\end{align} 
and noting that the energy density of our universe is presently $\rho \sim {\rm meV^4}$. This remarkable result can be understood as a consequence of our universe being both flat and critical, to within current measurement precision \cite{Aghanim:2018eyx}. In other words, the vacuum dominated energy density we have observed within a Hubble radius is very near to the energy density required to form a black hole.

On the other hand, using the Hubble parameter as IR cutoff and treating Eq.~\eqref{eq:ineqorig} as an equality,
\begin{align}
\delta \rho_\Lambda = \frac{3 }{8 \pi} H^2 M_P^2.
\label{eq:hubprox}
\end{align}  
immediately predicts a problematic equation of state for vacuum energy in our universe \cite{Hsu:2004ri}. Our universe was initially radiation and then matter dominated. During these expansionary epochs, $H$ will evolve with the expansion scale factor $a$ like $1/a^{2}$ and $1/a^{3/2}$, respectively. This implies that vacuum energy $\delta \rho_\Lambda$ in Equation \eqref{eq:hubprox} will also redshift approximately like matter and radiation during these epochs. Reference \cite{Hsu:2004ri} pointed out that Eq.~\eqref{eq:ineqorig} implies a non-standard equation of state, specifically $w \equiv \frac{\delta P}{\delta \rho} > 0$, for the vacuum energy density $\delta \rho_\Lambda$. This can be excluded using present measurements of the equation of state for dark energy  $w = -1.03 \pm 0.03$ \cite{Aghanim:2018eyx}. Subsequently, Reference \cite{Li:2004rb} proposed using the future event horizon of the universe as the IR cutoff in \eqref{eq:ineqorig}, $L = a \int^{\infty}_t \frac{dt}{a}$. However, the most straightforward version of this future horizon proposal predicted $w \approx -0.9$.

In the remainder of this article we will advocate for a different treatment than CKN for setting a black hole bound on the validity of effective field theories. While CKN considered only virtual ($i.e.$ loop-contributions) to vacuum energy density, we propose that the correct energy density threshold at which effective field theory breaks down is determined by both physical and virtual energy densities. In contrast to CKN, which predicted $\delta \rho_\Lambda \sim {\rm meV^4}$, our proposal leads to the conclusion that loop-contributions to vacuum energy can be consistently set to zero, $\delta \rho_\Lambda \sim 0$.  In the remainder of this paper we detail this ``matter effect" variant of the CKN solution to the cosmological constant problem, which is consistent with the observed equation of state for dark energy ($w=-1$), implies certain bounds on the curvature of the universe, and predicts new corrections to lepton $g-2$ measurements at futuristically precise experiments.

Before continuing, for the sake of clarity it is worth pointing out that we will exclusively treat the dark energy that has been observed in our universe as either a constant vacuum energy or a (not necessarily static) vacuum energy provided by loop corrections in what follows. We include a brief discussion of the implications of quintessence fields and other time-varying sources of dark energy in our conclusions.

\section{Matter Effect For Gravitational UV/IR Mixing}
\label{sec:mattereffect}

Incorporating a matter effect substantially changes the gravitational UV/IR mixing solution to the cosmological constant problem. While the UV/IR mixing contribution to the vacuum energy of the universe proposed by CKN requires that zero-point loop corrections to the vacuum energy density not exceed the threshold where a black hole would form in a space of size $L$, we propose that the restriction should be extended to include any matter, radiation, and ``bare'' vacuum energy contained in a space of size $L$. This proposal aims to incorporate the effect of physical field densities present in a system.

Our contention is that, insofar as a quantum field theoretic description of the volume enclosed in $L$ is breaking down when the field density of the system reaches a certain threshold, one should treat ``virtual'' and ``physical'' field configurations on equal footing. In other words, the new ``matter effect'' condition is that \emph{both} virtual contributions to the energy density of a system \emph{and} physical or ``on-shell'' field configurations, should not exceed the energy density of a black hole filling the space occupied by the system,
\begin{align}\label{eq:RschwME}
L \gtrsim R_{Schw} = 2G (\delta \rho_\Lambda + \rho_0) \frac{4 \pi}{3} L^3 
\end{align}
where $\rho_0 = \rho_m + \rho_r + \rho_\Lambda $ is the sum of all physical contributions, matter, radiation, and vacuum energy, to the energy density of the system. Note especially that $\rho_0$ includes any physical ($\rho_\Lambda$) as opposed to virtual ($\delta \rho_\Lambda$) contribution to the vacuum energy.

The matter effect gravitational UV/IR inequality indicating the breakdown of effective field theory for calculating loop contributions to vacuum energy is
\begin{align}
\label{eq:matineq}
\delta \rho_\Lambda + \rho_0 & \lesssim  \frac{3}{8 \pi} \frac{M_{P}^2 }{ L^2}.
\end{align}
Setting this to an equality leads to a vacuum energy contribution
\begin{align}
\label{eq:mateq}
\delta \rho_\Lambda =  \kappa  \frac{3}{8 \pi} \frac{M_{P}^2 }{ L^2} - \rho_0.
\end{align}
In the final expression we have introduced an order unity constant $\kappa$, which parameterizes the equality. In the case that effective field theory breaks down when field energy densities imply the formation of an uncharged, spin zero black hole, we expect $\kappa = 1$.

\section{Matter Effects And UV/IR Mixing Cosmology}
\label{sec:cosmo}

In order to apply the matter effect UV/IR relation to any system, including the observable universe, we need to determine the length scale which defines the IR cutoff of the system, $\Lambda_{IR} = 1/L$. The Hubble horizon has been previously identified as an appropriate IR cutoff for UV/IR relations in \cite{Cohen:1998zx}, $L^{-1}=H = \Lambda_{IR}$. It defines the boundary beyond which material recedes from an observer faster than the speed of light, and so it provides a well-motivated choice for the IR cutoff of the universe, since we are interested in defining a boundary within which field densities are restricted by the density of a black hole contained within the same radius. In contrast, \cite{Li:2004rb} proposed using the proper distance to the event horizon of the universe as an IR cutoff, $L= a \int_0^\infty \frac{dt}{a} = a \int_a^\infty \frac{da}{H a^2 }$, which resulted in an equation of state for dark energy closer to $w \approx -0.9$. In the following treatment of matter effect UV/IR mixing, we will use the Hubble horizon as the relevant IR cutoff length scale. For a discussion of other IR cutoff choices, see Section \ref{app}.

In the case that our length scale is given by the Hubble horizon, Eq. \ref{eq:mateq} becomes
\begin{align}\label{eq:eqhubb}
\rho_r + \rho_m + \rho_{\Lambda} + \delta \rho_\Lambda &= \kappa \frac{3}{8 \pi} M_{P}^2 H^2.
\end{align}
This equation bears a striking resemblance to the Friedmann equation in flat space, $3 M_{P}^2 H^2/8 \pi = \rho_0 + \delta \rho_{\Lambda} $, although it has been derived by other means, namely by placing black hole density restrictions on the validity of effective field theory within a Hubble radius. Note in particular that for a flat universe and $\kappa = 1$, loop corrections to the cosmological constant will vanish, $\delta \rho_{\Lambda} \rightarrow 0$, if the vacuum energy in our universe is attributed to a cosmological constant $\rho_\Lambda = V_{obs}^4$. In this case, the Friedmann equation becomes
\begin{align}
\frac{3}{8 \pi} M_{P}^2 H^2 = \rho_r + \rho_m + \rho_{\Lambda}.
\end{align}
Substituting this into Eq.~\eqref{eq:eqhubb} with $\kappa =1$, we find $\delta \rho_{\Lambda} = 0$. Evidently matter effect gravitational UV/IR mixing addresses the cosmological constant problem by predicting that effective field theory breaks down for any $\delta \rho_{\Lambda} > 0$, in the presence of a physical vacuum energy density, $\rho_\Lambda =V_{obs}^4$. This altogether implies a constant background vacuum energy $V_{obs}^4$, with effective field theory breaking down if any additional zero-point loop contributions to vacuum energy are introduced.

\subsection{Flat and Open Curvature in Gravitational UV/IR Mixing}

Matter effect gravitational UV/IR mixing makes some loose predictions regarding the curvature of the universe. To understand curvature in gravitational UV/IR mixing cosmology, we first consider the full Friedmann equation for a homogeneous, isotropic universe,
\begin{eqnarray}\label{eq:fr}
\frac{3}{8 \pi} H^2 M_P^2 = \rho_r + \rho_m + \rho_{\Lambda} + \delta\rho_\Lambda = \rho_k
\end{eqnarray}
where $\rho_k \equiv \frac{-k}{a^2}$ accounts for the curvature of the universe and $k=-1,0,+1$ for an open, closed, and flat universe, respectively.

Substituting this full Friedmann equation into Eq. \eqref{eq:mateq} with $L = H^{-1}$, we find
\begin{eqnarray}\label{eq:mainA}
\frac{3}{8 \pi} H^2 M_P^2 = \kappa \frac{3}{8 \pi} H^{2} M_P^2 + \rho_k
\end{eqnarray}
As we have already noted setting $\rho_k=0$ and $\kappa=1$, which is consistent with present measurements \cite{Aghanim:2018eyx}, implies a flat universe and is compatible with vanishing zero-point loop corrections to vacuum energy, and the result of that computation was the Friedmann equation for a flat universe.

We have seen that matter effect UV/IR mixing would be consistent with a perfectly flat universe. However, our starting point in this regard, Equation \eqref{eq:matineq}, really only gave an approximate prediction for the scale at which effective field theory breaks down based on a Schwarzschild black hole solution. Therefore, we will examine what happens when $\kappa$ and $\rho_k$ take on different values, assuming the Schwarzschild black hole UV/IR bound, Eq.~\eqref{eq:matineq}, applies in curved spacetimes. Using the Friedmann equation and setting $L=H^{-1}$ we can rephrase Eq. \eqref{eq:matineq} as
\begin{align} \label{eq:curve}
\delta \rho_\Lambda + \rho_0  \leq \kappa_d \frac{3}{8 \pi} \frac{M_{P}^2 }{ L^2}, \nonumber \\
\left( 1 - \kappa_d \right) \frac{3}{8 \pi} H^2 M_P^2  \leq \rho_k,
\end{align}
where $\kappa_d$ now defines deviations from the inequality given in Eq.~\eqref{eq:matineq}. We recall that around Eqs.~\eqref{eq:matineq} and \eqref{eq:mateq} we argued that our effective field theory is supposed to break down for $\kappa_d \gtrsim 1 $, since this implies a region inside radius $H^{-1}$ with a total energy density greater than that required to form a black hole. However, we recall that the inequality given in Eq.~\eqref{eq:matineq} used the Schwarzschild horizon as a limit on effective field theories contained in some radius $L$, but this would seem to depend on the details of how field theories are altered at this limit. However, as we will see Eq.~\eqref{eq:curve} implies some cosmological restrictions on the range of values $\kappa_d$ can take, based on the non-observation of curvature in our universe. 

To see how these restrictions arise, we first suppose that $\kappa_{d} \leq 1$, which implies that effective field theories are altered a bit before the Eq.~\eqref{eq:matineq} limit on field theories is reached. In this case $\rho_k$ will never be negative, which implies that the universe is either flat or open. But in fact, for $\kappa_d < 1$, the universe is necessarily open. Moreover, the observation of a nearly flat universe thus requires that $\kappa_d$ is not too small, $\kappa_d \geq 0.994$ at $2 \sigma$, using both cosmic microwave background and baryon acoustic oscillation data \cite{Aghanim:2018eyx}.

\subsection{Closed Universe Limitations}

It is also possible to fruitfully consider the case that $\kappa_d > 1$, which implies that the inequality in Eq.~\eqref{eq:matineq} is slightly violated. For the case that $\kappa_d > 1$, the matter effect UV/IR mixing formula would predict a closed universe. It also implies a limit on how large curvature will become in the future. First, we note that the only non-constant part of the left hand side of Eq. \eqref{eq:curve} is $H^2$. This means that we can define a constant 
\begin{eqnarray}
- C_0 \equiv \left(1 -  \kappa_d \right) \frac{3}{8 \pi} M_P^2,
\end{eqnarray}
where this definition emphasizes that $\kappa_d > 1$. Eq. \eqref{eq:curve} becomes 
\begin{eqnarray}
- C_0 \dot{a}^2 a^{-2} = - \left|\rho_{k}^{\left(a=1\right)}\right| a^{-2},
\end{eqnarray}
where $\rho_k$ has been defined at a reference scale ($a=1$). This yields a minimum value for $\left|\dot{a}\right|$,
\begin{eqnarray}\label{eq:negk}
\dot{a}^2 \geq \left|\frac{\rho_{k}^{\left(a=1\right)}}{C_0}\right|.
\end{eqnarray}
We see that since $\dot{a}$ is positive at the present time, it will always be positive, which implies that a matter effect UV/IR closed universe cannot have both finite space and finite time.

We now obtain a restriction on the curvature of this theory using the Friedmann equation,
\begin{align}\label{eq:fried2}
\frac{3}{8 \pi} M_P^2 \frac{\dot{a}^2}{a^{2}} &= a^{-4} \rho_{r}^{\left(a=1\right)} + a^{-3} \rho_{m}^{\left(a=1\right)} \nonumber \\ &+ a^{-2} \rho_{k}^{\left(a=1\right)} + \rho_{\Lambda}^{\left(a=1\right)} + a^{-3-3w} \delta\rho_{\Lambda}^{\left(a=1\right)},
\end{align}
where we have defined all energy densities as being fixed at reference scale $a=1$. In what follows we will also define $C_F \equiv \frac{3}{8 \pi}  M_P^2$ for convenience. Now schematically, we know that our universe is radiation dominated at early times.
Therefore, there is a value of $\dot{a}$ at early times where the radiation-dominated energy density of the universe is approximately $ \rho_0$ and
\begin{align}
\dot{a}^2 = - \left|\frac{\rho_{k}^{\left(a=1\right)}}{C_F}\right| + \frac{\rho_0}{C_F a^2} \geq \left|\frac{\rho_{k}^{\left(a=1\right)}}{C_0}\right| = \left|\frac{\rho_{k}^{\left(a=1\right)}}{C_F \left( \kappa_d - 1 \right)}\right|,
\end{align}
where the inequality on the right hand side of this expression is obtained from Eq.~\eqref{eq:negk}. This in turn implies a bound on the curvature of a closed UV/IR mixed universe,
\begin{eqnarray}
\frac{\rho_0}{\left|\rho_{k}\right|} \geq  \frac{\kappa_d}{\kappa_d - 1}.
\end{eqnarray}

\section{Alternative IR Cutoffs}
\label{app}
For a cosmological system like our universe, one might consider choosing one of three seemingly relevant length scales to set the IR cutoff -- the particle horizon which determines what could have affected us in the past, the Hubble horizon which is the standard relevant length, and the event horizon, which indicates the future causally connected volume. In the limit that the Hubble horizon is constant, all three scales are the same. See \cite{Li:2004rb,Hellerman:2001yi,Banks:2001px} for examples of these horizons being utilized.

In preceeding derivations, we have assumed that the IR cutoff length $L$ is given by the Hubble horizon $\frac{1}{H}$. Now we will examine the case when $L$ is given by two other horizons, the particle horizon
\begin{align}
L_p = a \int^t_0 \frac{dt}{a},
\end{align}
and the event horizon
\begin{align}
L_e = a \int^{\infty}_t \frac{dt}{a}.
\end{align}
Neither will prove useful for the purposes of constructing a realistic cosmology based on matter effect gravitational UV/IR mixing. Starting with the matter effect UV/IR inequality \eqref{eq:curve},
\begin{align}\label{eq:nonhl}
\frac{3}{8 \pi} H^2 M_P^2 \leq \kappa \frac{3M_{P}^2 }{8 \pi L^2}.
\end{align}
If we choose $\kappa=1$ as above, it immediately follows that
\begin{align}\label{eq:lhrel}
L \leq \frac{1}{H}.
\end{align}
This implies a Hubble horizon that is larger than or equal to the particle or event horizon. Neither case is permitted in our universe, where it has been observed that $L_p > H^{-1}$ and $L_e > H^{-1}$ \cite{Davis:2003ad}.

\section{Experimental Consequences For Matter Effect UV/IR Mixing}

From Eq.~\eqref{eq:mateq}, it follows directly that matter effect gravitational UV/IR mixing predicts observable consequences for particle experiments with sensitivity to new dynamics at an effective field theory cutoff
\begin{eqnarray}
\Lambda_{UV}^4 \simeq \frac{M_P^2}{2 L^2 } - \rho_0,
\label{eq:uvscale}
\end{eqnarray}
where we reiterate that $\rho_0$ is the energy density in the region that the experimental measurements take place and $L$ is the length scale of the region, which is related to the infrared cutoff as $\Lambda_{IR} \sim \frac{1}{L} $. Note that unlike Eq.~\eqref{eq:mateq}, Eq.~\eqref{eq:uvscale} has not assumed a spherical geometry for the region bounded by $L$.

In the original gravitational UV/IR proposal, CKN noted that electron $g-2$ measurements might some day reach the precision required to observe non-local gravitational corrections. CKN calculated the minimum possible correction to the electron $g-2$ by simultaneously varying both UV and IR cutoffs \cite{Cohen:1998zx}. On the other hand, in a companion paper \cite{BG2}, we have found that a straightforward choice for the infrared cutoff $\Lambda_{IR}$, leads to interesting and measurable consequences for the muon $g-2$. Indeed, we have found that this could serve as an explanation of the anomalous muon $g-2$ anomaly observed in the last decade. Thus far, the muon g-2 has been measured to match Standard Model predictions at parts-per-billion precision. The Brookhaven Muon E821 measurement \cite{Bennett:2006fi} of \mbox{$a_\mu^{exp} \equiv (g-2)_\mu/2 =11 659 208.0(6.2) \times 10^{-10}$} can be compared to the Standard Model prediction \cite{Keshavarzi:2018mgv}, \mbox{$a_\mu^{SM} = 11 659 182.05 (3.56)\times 10^{-10}$}.

However, for the purposes of calculating the relative impact of a matter effect UV/IR mixing correction to lepton $g-2$ measurements, $\delta(g-2)_{ME}$, as compared to the original UV/IR mixing correction $\delta(g-2)_{UV/IR}$, here we will not need to extensively address the definition of the IR cutoff $\Lambda_{IR}$. 

With the UV cutoff for new dynamics defined as in Eq.~\eqref{eq:uvscale}, we would like to consider whether terrestrial experiments could reach the precision necessary to test matter effect UV/IR mixing. The leading order muon g-2 correction from UV/IR mixing is \cite{Peskin:1995ev}
\begin{align}
\delta(g-2) \simeq \frac{\alpha}{\pi}\left( \frac{m_\mu}{\Lambda_{UV}} \right)^2  &=  \frac{\alpha}{\pi}\left( \frac{m_\mu^2}{\sqrt{\frac{ M_P^2}{2 L^2 } - \rho_0}} \right) 
\nonumber
\\ & = \sqrt{2} \frac{ \, \alpha m_\mu^2 L}{\pi M_P}\left(1 + \frac{L^2 }{M_P^2}\rho_0 + \cdots \right),
\label{eq:mec}
\end{align}
where $\alpha$ is the fine structure constant and we have expanded in the limit that the matter effect correction is small, $\rho_0 \ll \frac{M_P^2}{2L^2 }$, which we believe is satisfied by any presently conceivable terrestrial experiment. 
Therefore, the matter effect correction to UV/IR mixing in the laboratory, relative to the non-matter effect UV/IR mixing correction ($i.e.$ the first expansion term relative to the second in Eq.~\eqref{eq:mec}), is approximately
\begin{align}
\delta(g-2)_{ME} - &\delta(g-2)_{UV/IR}  \simeq  \frac{L^2 }{M_P^2}\rho_0 \nonumber \\ & \simeq 10^{-13} \left( \frac{\rho_0}{ {\rm g/cm^3}} \right)\left( \frac{L}{10^7~{\rm cm}} \right)^2,
\end{align}
where we have adopted a terrestrial density of gram/${\rm cm^3}$ and a plausible collider length scale of $L=10^7$ cm. We see that the precision measurements of lepton magnetic moment required to test matter effect gravitational UV/IR mixing lie beyond present experimental capabilities. In particular, the precision of the Brookhaven E821 muon g-2 measurement would need to be improved by over ten orders of magnitude for the matter effect to become relevant.  However, this matter effect would be sought after first finding a larger vacuum gravitational UV/IR mixing effect, which could be found at current experiments \cite{BG2}. 

\section{Conclusions}
\label{conclude}

We have introduced a matter effect variant of the gravitational UV/IR mixing prescription for the cosmological constant problem. It appears that non-local gravitational effects could indeed resolve the cosmological constant problem, so long as physical field densities are properly incorporated when determining the UV cutoff for which field theory breaks down in a space of size $L$. As we have discussed, UV/IR mixing appears to mildly prefer a flat or open universe, since in these cases the black hole density bound is not saturated. However, slightly beyond the naive limits of the bound, a closed universe with a restricted curvature is also possible. The matter effect UV/IR theory predicts local-density-dependent corrections to extremely precise measurements of fundamental constants, like the magnetic moment of electrons and muons.

Many related topics remain open for further exploration. While we have shown that zero-point loop corrections to vacuum energy can vanish for a universe with a cosmological constant, which in this case means a constant background vacuum energy density, it would be interesting to determine how this proposal changes in the presence of a quintessence field, or some other source of varying vacuum energy density. Along similar lines, future work might also consider whether zero-point loop corrections could account for a subdominant portion of the observed vacuum energy, so that the left hand side of Equation \eqref{eq:mateq} is nonvanishing. Such a proposal could be constrained by bounds on the equation of state for dark energy, which similarly restricted the original CKN proposal \cite{Cohen:1998zx}, which attributed all of the observed vacuum energy to zero-point loop corrections \cite{Hsu:2004ri,Li:2004rb}.

We have found the implications of a gravitational UV/IR mixing matter effect for particle experiments, but we have assumed a constant background matter density. A time-varying background matter density should also be considered. We look forward to exploring these and other aspects of gravitational UV/IR mixing in future work.

\section*{Acknowledgements}
We wish to thank the anonymous referees for many helpful comments and suggestions. We wish to particularly thank one anonymous referee for a discussion of the possibility that a subdominant portion of the observed vacuum energy could be sourced by zero-point loop corrections. We acknowledge the support of the Natural Sciences and Engineering Research Council of Canada. Research at Perimeter Institute is supported by the Government of Canada through Industry Canada and by the Province of Ontario through the Ministry of Economic Development \& Innovation. 
\appendix

\bibliography{uvirmatter}

\end{document}